# Alkaline-Earth Rare-Earth Fluoride Nanoparticle Superlattices for Ultrafast, Radiation Stable Scintillators

Parivash Moradifar[1*+], Tim Brandt van Driel[2*], Masashi Fukuhara[1], Cindy Shi[1], Ariel Stiber[1], Federico Moretti[3], Qingyuan Fan[1,4], Diana Jeong[5], Aaron M. Lindenberg[1,4], Garry Chinn[5], Craig S. Levin[5+], Jennifer A. Dionne[1,5+]

1 Department of Materials Science and Engineering, Stanford University, Stanford, CA, USA
2 Linac Coherent Light Source, SLAC National Accelerator Laboratory, Menlo Park, CA, USA
3 Lawrence Berkeley National Laboratory, Cyclotron Road, Berkeley, CA, USA
4 Stanford PULSE Institute, SLAC National Accelerator Laboratory, Menlo Park, CA, USA
5 Department of Radiology, Stanford University, Stanford, CA, USA

*These authors contributed equally to this work.
+Correspondence and requests for material should be addressed to Parivash Moradifar (email:pmoradi@stanford.edu), Craig S. Levin (cslevin@stanford.edu), Jennifer Dionne (email: jdionne@stanford.edu).

**Abstract**

Radioluminescent nanostructures provide a pathway to the fabrication of next-generation scintillators with tunability in composition, size, and morphology, and spectral and temporal properties, as well as scalable processing. Here we create a 3D millimeter-scale solid-state scintillators from SrLuF:$Ce^{3+}$/$Pr^{3+}$@SrLuF core-shell nanostructures, integrating nanoscale building blocks into self-assembled macroscopic crystals. These scintillators exhibit single-digit nanosecond decay times, linear response, resistance to radiation-induced degradation, and optical emission yields within an order of magnitude of YAG:$Ce^{3+}$. We select a SrLuF host lattice owing to its high effective atomic number, wide band gap, and low phonon energy, which together support efficient 4f-5d radiative transitions from $Ce^{3+}$ and $Pr^{3+}$ activators while suppressing afterglow. We create a library of core-shell nanoscintillators with undoped SrLuF shells and cores spanning compositions from undoped SrLuF to fully doped SrCeF or SrPrF with the enhanced luminescence observed at high dopant concentrations. Time-resolved and steady-state X-ray excited optical luminescence (XEOL) reveal broadband emission at 310 nm ($Ce^{3+}$) and 335 nm ($Pr^{3+}$) with biexponential decays in the sub-nanosecond (~100-500 ps) and sub-15 ns (4-13 ns) regimes, demonstrating tunable radiative efficiency and ultrafast dynamics. Ensemble performance of the mm-scale superlattices is characterized under both continuous-wave and femtosecond high-intensity excitation, revealing high light yield, linear response, and radiation hardness under extreme irradiation of ultrafast 50fs X-ray pulses up to 5mJ/mm$^2$ corresponding to a peak intensity of $10^{13}$ W/cm$^2$. Together, these results establish a design framework for stable, bright, and tunable scintillation platforms with applications in precision health, space exploration, real-time radioactive waste monitoring, and hard X-ray imaging at next-generation free-electron laser facilities.

**Introduction**

For nearly a century, scintillators have transformed our ability to see and sense high-energy radiation.[1–3] They remain indispensable in diverse technologies spanning medical imaging (positron emission tomography, single photon computed tomography, X-ray computed tomography),[4,5] homeland security (radiation monitoring and nuclear threat detection),[6,7] high-energy physics (particle tracking and calorimetry),[8] space exploration (cosmic-ray and gamma-ray telescopes),[9] and environmental monitoring (dosimetry and contamination assessment).[10,11] Scintillators convert ionizing radiation (α, β, γ, X-rays, electrons, neutrons) into ultraviolet or visible photons, acting as wavelength shifters that make invisible radiation detectable. Following Röntgen's discovery of X-rays, calcium tungstate ($CaWO_4$) and silver-doped zinc sulfide (Ag-ZnS) became the first scintillators, with $CaWO_4$ enabling X-ray phosphors for imaging the human skeleton as early as 1896.[12–14]

The scintillation process involves three sequential stages: (i) creation, multiplication, and thermalization of charge carriers, (ii) transport of electronic excitation energy through the host lattice, including carrier- and excitonic-mediated processes, and (iii) radiative recombination at luminescent centers, such as activator emission or intrinsic host channels including cross-luminescence. Each stage operates on timescales from picoseconds (thermalization) to milliseconds (luminescence emission).[15] Host composition, defect density, and the presence of dopants are among parameters that govern how efficiently the absorbed radiation energy is converted into photons, and thus define a scintillator's timing resolution, light yield, and radiation hardness.[16–18]

Inorganic scintillators are broadly categorized as intrinsic or activated. Intrinsic materials emit directly from their host lattice via excitonic processes,[19] whereas activated scintillators rely on dopants such as $6s^2$ ions ($Tl^+$, $Pb^{2+}$, $Bi^{3+}$) or lanthanides ($Ce^{3+}$, $Eu^{3+}$) to establish efficient luminescent centers.[20] The host lattice strongly influences these processes and interaction characteristics: halides (NaI, CsI) typically exhibit strong γ- and X-ray attenuation arising from their high effective atomic number and density, oxides (BGO, LSO) offer structural and radiation robustness, sulfides enable strong absorption through smaller band gaps, and fluorides combine wide band gaps with low phonon energies, supporting efficient radiative transitions and high energy resolution. Low-phonon-energy host lattices suppress multiphonon nonradiative relaxation, thereby preserving excitation energy for radiative recombination at activator centers, leading to enhanced light yield, reduced thermal quenching, and improved scintillation efficiency.

Despite decades of progress, conventional scintillators face intrinsic limitations: low light-extraction efficiency, restricted emission tunability, slow decay dynamics, and fabrication constraints requiring high-temperature solid-state synthesis (>1000 °C).[2,19] As a result, most commercial scintillators operate at efficiencies below 10% and lack the tunability and ultrafast response needed for next-generation applications in real-time imaging, low-dose diagnostics, and high-rate particle detection. Radioluminescent nanostructures, or nanoscintillators, provide a promising solution. By engineering composition, morphology, and anisotropy at the nanoscale, nanoscintillators overcome many of the trade-offs inherent to bulk materials.[20–22] Their high surface-to-volume ratios, flexible core-shell architectures, and compatibility with colloidal synthesis enable precise control over emission wavelength, radiative dynamics, and energy transfer pathways. Among available hosts, alkaline-earth rare-earth fluorides ($M_{1-x}Ln_xF_{2+x}$, M = Sr, Ba, Ca) stand out for their chemical stability, high rare-earth solubility, and photostability.[23] They are already proven platforms for photon upconversion, where lanthanide doping allows efficient NIR-to-UV/visible conversion for applications in sensing, photocatalysis, and bioimaging.[24–27] Leveraging the outstanding photophysical properties for down-conversion offers an avenue to build scintillators that combine environmental stability and high stopping power with ultrafast optical response.

Here, we introduce a new class of freestanding, millimeter-scale 3D solid-state superlattice scintillators constructed from SrLuF-based core-shell nanoscintillating building blocks doped with varying concentrations of $Ce^{3+}$ and $Pr^{3+}$ to optimize brightness and lifetime at the single building block scale. The SrLuF host provides high density, high effective atomic number (~54.5), and low phonon energy, while the epitaxial shell minimizes defect formation and stabilizes the dopant environment. $Ce^{3+}$ contributes dipole allowed 5d-4f transitions with reduced afterglow, whereas $Pr^{3+}$ offers dual inter- and intra-configurational emissions with ultrafast decay channels. Each superlattice is assembled from compositionally homogeneous nanocrystals containing a single dopant species at a fixed concentration. Guided by X-ray excited optical luminescence (XEOL) characterization across the nanocrystal library, we identify SrCeF@SrLuF and SrPrF@SrLuF as optimal candidates for maximizing brightness and ultrafast response, respectively, and use these compositions to fabricate transparent 3D superlattices for testing under ultrashort, high-fluence X-ray excitation at an XFEL. By scaling these nanoscale building blocks into transparent 3D superlattices, we demonstrate scintillation performance approaching that of commercial YAG:$Ce^{3+}$ within an order of magnitude in light output under extreme irradiation using ultrashort and ultrabright X-ray sources, while simultaneously enabling X-ray object imaging using the superlattices as scintillation screens. Our work introduces a scalable design framework for high-performance nanoscintillators and paves the way for next-generation applications in precision health, high-repetition-rate XFEL imaging, space exploration, and radiation monitoring in extreme environments.

**Results and Discussions**

We create an extended library of nanoscintillators by systematically substituting $Lu^{3+}$ sites in SrLuF with $Ce^{3+}$ and $Pr^{3+}$, yielding monodisperse, sub-20 nm MLnF nanoparticles with controlled dopant concentrations spanning from undoped (0%) to fully doped (100%). To demonstrate the scalability and flexibility of the nanoscintillator design, we explore two prototypes of highly transparent mm-scale scintillators: (i) flexible nanocomposite scintillators and (ii) self-assembled scintillators, a bulk-size prototype where the nanoscintillators serve as the fundamental building blocks. The flexible scintillator film is based on a polydimethylsiloxane (PDMS) matrix (Figure S1) with SrLuF nanoscintillators embedded to form a nanocomposite, while the self-assembled scintillator is prepared by controlled solvent evaporation.

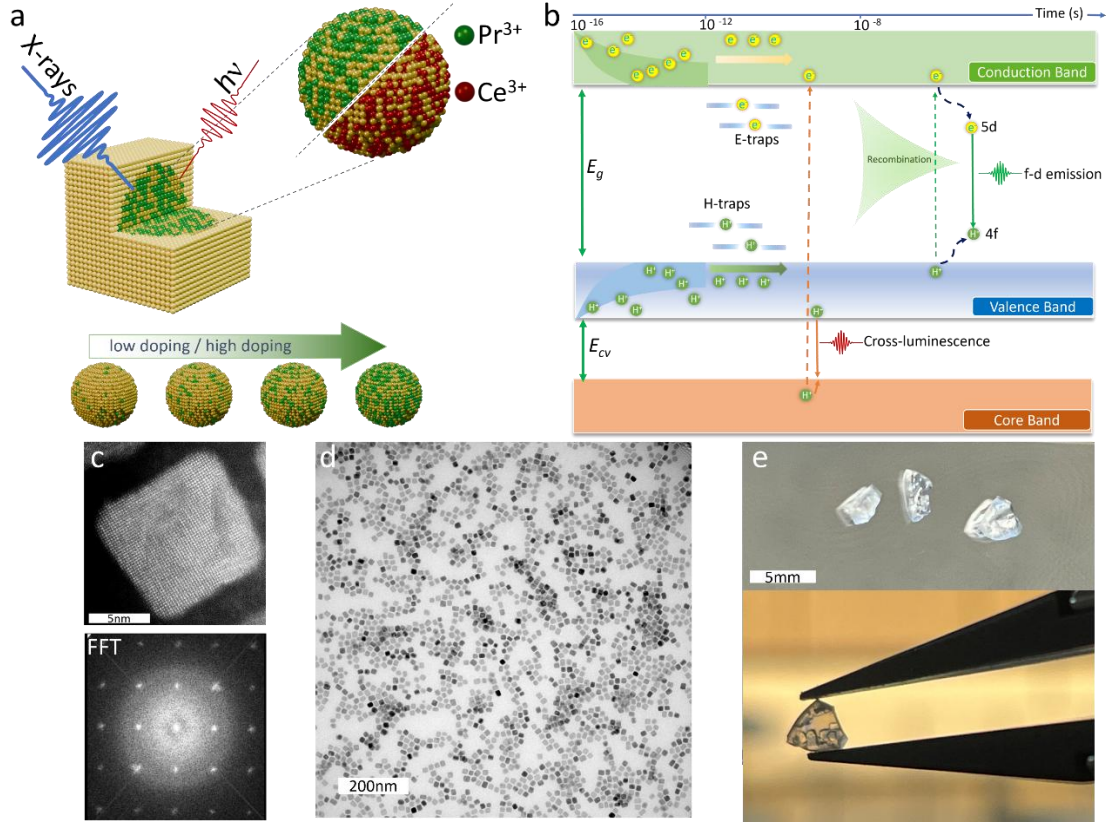

**Figure 1: SrLuF core-shell nanoscintillator building blocks and scalable assemblies.**
(a) Schematic of the SrLuF:RE$^{3+}$@SrLuF core-shell nanoscintillator showing a tunable doped core (RE = Ce$^{3+}$, Pr$^{3+}$) and an undoped shell.
(b) Schematic of the scintillation mechanisms showcasing two radiative decay pathways in SrLuF -based nanoscintillators: dipole-allowed 5d → 4f transitions associated with the Ce$^{3+}$/Pr$^{3+}$ activators, and cross-luminescence (CL) arising from valence-to-core recombination in the SrLuF host.
(c) Representative TEM image of a single nanocube building block (upper) with corresponding Fast Fourier Transform (FFT) (lower) confirming single-crystal FCC crystal structure.
(d) Representative TEM image showing nanoparticle morphology and size distribution.
(e) Optical images of mm-scale optically transparent crystals produced by controlled self-assembly of millions of nanocubes (see SI for the PDMS nanocomposite prototype).

We construct $M_{1-x}Ln_xF_{2+x}$ core–shell nanoscintillators comprising a doped core encapsulated by an undoped SrLuF shell with a final nanocube morphology (Figure 1a). Radiative response in Ce$^{3+}$- and Pr$^{3+}$-containing nanostructures proceeds through two channels: dipole-allowed 5d→4f transitions of the activator ions that generate ultrafast UV-visible emission, and a potential host-mediated cross-luminescence (CL) arising from valence-to-core recombination intrinsic to wide-bandgap SrLuF (Figure 1b). Cross-luminescence as an intrinsic emission pathway occurs even in undoped nanoparticles and provides an additional ultrafast emission channel that can couple to dopant ions in $Ce^{3+}/Pr^{3+}$ samples, collectively contributing to tunable radiative decay dynamics. These two pathways, dipole-allowed 5d→4f transitions and cross-luminescence, establish the foundation for the scintillation behavior described hereafter (e.g., in Figure 3). We verify structural homogeneity at the nanoscale through single-particle transmission electron microscopy (TEM) and Fast Fourier ransform (FFT) analysis, confirming crystallinity and FCC symmetry (Figure 1c and Figure S2), and demonstrate morphological uniformity across nanoparticle ensembles (Figure 1d). Directed assembly of these nanocubes yields 3D millimeter-scale transparent crystals (Figure 1e), bridging nanoscale photophysical control and macroscopic scintillator architectures.

Our scalable solution process, based on the thermal decomposition of alkaline-earth (M) and rare-earth (Ln) trifluoroacetate salts in a high-boiling solvent under inert conditions, produces uniform spherical seed nanoparticles with surface nucleation sites. The core nanoparticles serve as templates for the epitaxial growth of an undoped passivation shell via a dropwise hot-injection technique, resulting in a transition to nanocube morphology.[28,29] The undoped shell suppresses surface-related nonradiative recombination pathways and reduces defect-mediated quenching associated with surface and near-surface defect populations, including Frenkel defects, point-defect induced trap states (e.g. vacancy and vacancy-complex states) as well as undercoordinated surface sites, thereby enhancing luminescence efficiency.

Our library of core-shell nanoscintillator building blocks with varying dopant type and concentration provides a platform for systematically investigating correlated photophysical properties. TEM images confirm a narrow size distribution with consistent nanocube morphology across all compositions independent of dopant type or concentration as shown in Figure 2a. Figure 2b and 2c capture the transition from spherical core nanoparticles to the final nanocube building blocks after shell growth. Figure 2d shows a TEM image of densely packed nanoparticles dropcast on a TEM grid with the corresponding electron diffraction pattern; the ring pattern consisting of uniform intensity rings further confirms the crystal structure and uniform size distribution. Superlattice formation and structural ordering is inferred from electron microscopy and the emergence of macroscopic optical transparency, consistent with dense and uniform packing. SEM images of a mm-scale solid-state scintillator are shown in Figure 2e. These images and the inset X-ray diffraction pattern confirm long range assembly of the MLnF building blocks (note that the insulating nature of the nanocrystal building blocks and the surface charges challenge direct imaging of the superlattice structure). In particular, the X-ray diffraction pattern is consistent with the FCC electron diffraction from a 10 μm selected area aperture. This confirms that the sample retained its crystallinity and structure after the self-assembly, at the mm-length scale, and during measurements using an intense single-shot X-ray beam.

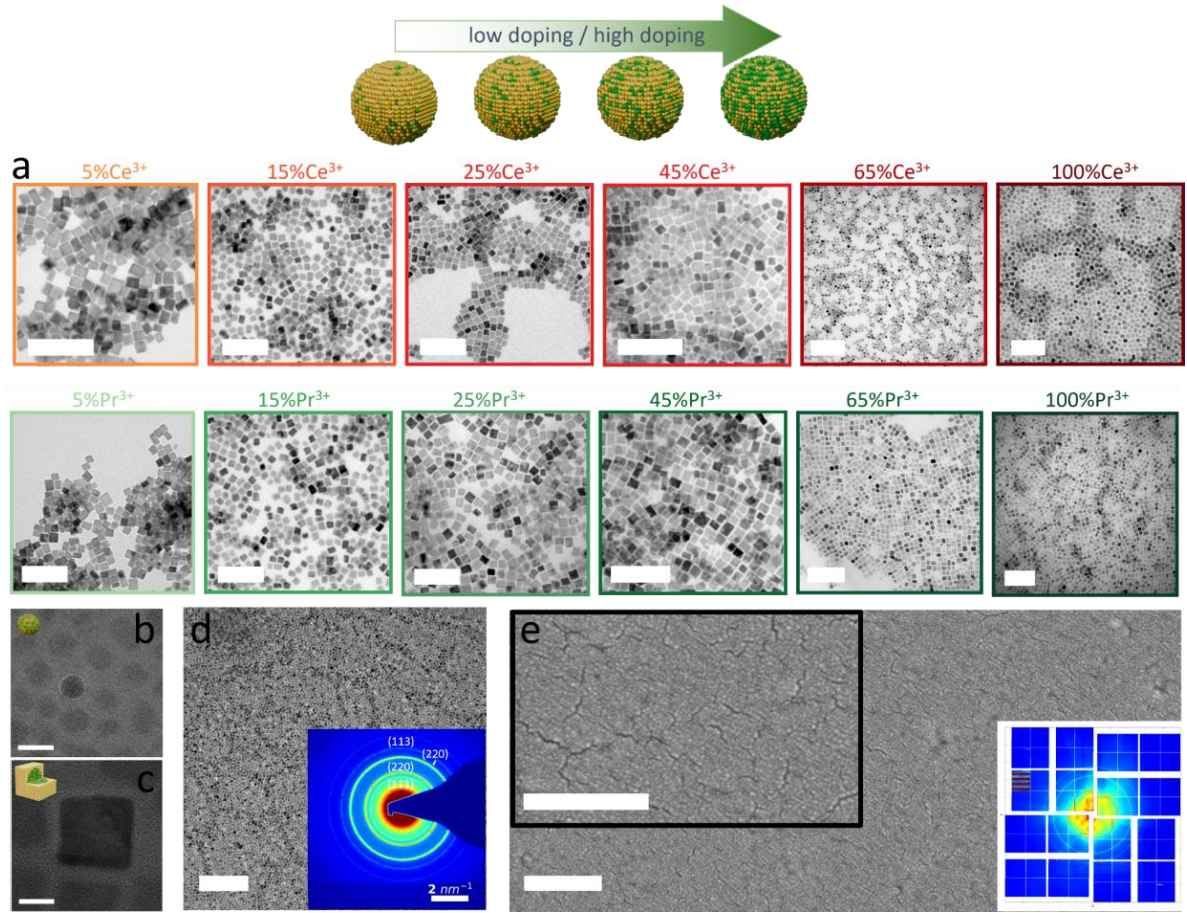

**Figure 2: Near-atomic engineering of SrLuF nanoscintillator nanocubes yields uniform morphology and structurally coherent macroscopic assemblies.**
(a) TEM images of monodisperse sub-20 nm SrLuF nanocubes across all dopant compositions, showing uniform morphology and narrow size distribution. (Scale bar: 100 nm.)
(b) TEM image of spherical core nanoparticles prior to shell growth. (Scale bar: 10 nm.)
(c) TEM image of the resulting epitaxial core–shell nanocubes after shell growth. (Scale bar: 10 nm.)
(d) TEM image of densely packed nanocubes and the corresponding selected-area electron diffraction pattern (inset). The ring sequence matches an FCC fluorite-type lattice with cations on an FCC sublattice and fluorine anions in tetrahedral sites. (Scale bar: 200 nm.)
(e) SEM image of the millimeter-scale self-assembled scintillator crystal showing long-range order, with accompanying single-shot ultrafast XRD (inset) consistent with the FCC reflections in (d). The low-Q background originates from air and substrate scattering.

It is worth noting that the low-Q background signal observed in the single shot X-ray diffraction pattern originates from air and substrates in the X-ray path. The combination of nanoscale control with scalable self-assembly into transparent crystals enables a versatile platform for designing ultrafast, high-efficiency scintillators. Further structural and compositional analyses are performed using X-ray diffraction (XRD), inductively coupled plasma optical emission spectroscopy (ICP-OES), and X-ray photoelectron spectroscopy (XPS). XRD measurements verify that all compositions, from undoped SrLuF to fully doped SrPrF and SrCeF, maintain a face-centered cubic (FCC) structure (space group Fm3̄m) with a consistent average lattice parameter of 5.7 Å, demonstrating that doping does not induce significant crystallographic distortions (Figure S3 and Table 1). ICP-OES measurements confirm that $Pr^{3+}$ concentrations closely follow nominal values across the entire doping range, while $Ce^{3+}$ concentrations match nominal values at low and high doping levels but show slight underdoping at intermediate concentrations (Figure S4). This deviation may influence the apparent concentration-dependent emission trends and should be considered when interpreting quenching behavior. XPS analysis of Ce 3d and Pr 3d spectra confirms both dopants are incorporated in the trivalent oxidation state, as evidenced by characteristic spin-orbit splitting: $Ce^{3+}$ binding energies at 879-890 eV ($3d_{5/2}$), 895-910 eV ($3d_{3/2}$), and $Pr^{3+}$ binding energy at 933.2 eV with a 20.5 eV spin-orbit splitting between $3d_{5/2}$ and $3d_{3/2}$ transitions (Figure S5).

To quantify stopping power, we calculated X-ray attenuation for 100 μm and 500 μm thick layers of SrLuF, SrCeF, SrPrF, and a mixed SrLuF:50%Ce assembly at 50% packing density and compared them with commercial YAG. Calculations employed tabulated mass attenuation coefficients from the NIST XCOM database to estimate transmission spectra (Figure S6). The results indicate effective stopping power exceeding that of YAG, particularly at higher photon energies (>25 keV). Pronounced absorption edges corresponding to the K-edges of Sr (16 keV), Ce (40.4 keV), Pr (42 keV), and Lu (63 keV) appear in the calculated spectra. Both thicknesses yield substantial attenuation of high-energy X-rays arising from the high-Z elemental composition, supporting the suitability of these nanoscintillator assemblies for high-energy radiation detection where most other nanoscintillator counterparts would fail otherwise.

The library of SrLuF nanoscintillators, featuring tunable dopant concentrations, enables a systematic investigation of how dopant type and concentration influence the scintillation characteristics in wide-bandgap fluoride hosts. To investigate the spectral and temporal characteristics of $Ce^{3+}$ and $Pr^{3+}$ doped SrLuF nanoscintillators and understand the compositional dependence of scintillation properties, we perform steady-state and time-resolved X-ray excited optical luminescence (XEOL) measurements across the entire library. An energy-level diagram summarizing the distinct 5d→4f and 4f→4f pathways of $Ce^{3+}$ and $Pr^{3+}$ is provided in Figure S7.
Figure 3a,b presents the time-integrated luminescence spectra for $Ce^{3+}$ and $Pr^{3+}$ doped nanoscintillators at room temperature respectively (details on setup and acquisition parameters are in the methods section). Under steady-state X-ray excitation at room temperature, $Ce^{3+}$-doped SrLuF nanoscintillators exhibit the expected broad parity allowed interconfigurational 5d→4f emission centered around 325 nm, with a subtle characteristic doublet structure corresponding to the $^2F_{5/2}$ and $^2F_{7/2}$ final states. The spectral width reflects strong electron-phonon coupling. In contrast, $Pr^{3+}$-doped SrLuF nanoscintillators exhibit a distinct spectral fingerprint. A series of sharp parity forbidden intraconfigurational $4f^2 \rightarrow 4f^2$ transitions appears in the visible range, including $^3P_0\rightarrow^3H_4$, $^3P_2\rightarrow^3H_5$, $^3P_1\rightarrow^3H_5$, $^1D_2\rightarrow^3H_4$, $^3P_0\rightarrow^3F_2$, and $^3P_0\rightarrow^3F_3$, superimposed on a broad band near

~310 nm. The sharp linewidths reflect minimal vibronic coupling in the shielded 4f orbitals, while the superimposed broadband component arises from parity-allowed $4f^1 5d^1 \rightarrow 4f^2$ emission.[30,31]

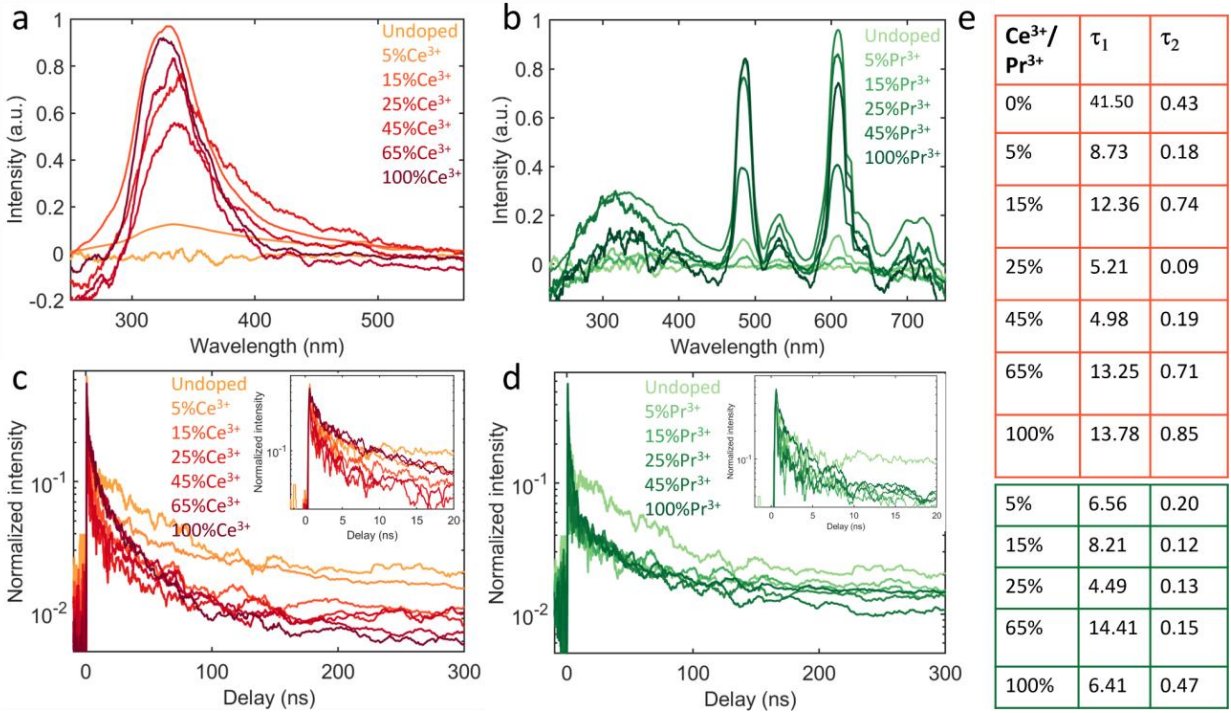

**Figure 3: Spectral and temporal scintillation characteristics of $Ce^{3+}$- and $Pr^{3+}$-doped SrLuF nanoscintillators.**

(a,b) Steady-state XEOL spectra of $Ce^{3+}$ and $Pr^{3+}$ doped SrLuF nanoscintillators at room temperature. $Ce^{3+}$ samples exhibit the characteristic broad 5d→4f emission centered near 325 nm with a $^2F_{5/2}$–$^2F_{7/2}$ doublet, while $Pr^{3+}$ samples display a mixed response consisting of a broadband ~310 nm component ($4f^1 5d^1 \rightarrow 4f^2$) superimposed with sharp $4f^2 \rightarrow 4f^2$ line emissions ($^3P_0 \rightarrow ^3H_4$, $^3P_2 \rightarrow ^3H_5$, $^3P_1 \rightarrow ^3H_5$, $^1D_2 \rightarrow ^3H_4$, $^3P_0 \rightarrow ^3F_2$, $^3P_0 \rightarrow ^3F_3$).

(c,d) Time-resolved XEOL decay curves for $Ce^{3+}$ and $Pr^{3+}$ doped samples showing biexponential behavior with sub-nanosecond and sub-15 ns components.

(e) Extracted radiative decay lifetimes from biexponential fits across dopant concentrations, highlighting the coexistence of dopant-centered 5d→4f ($Ce^{3+}$), 5d→4f ($Pr^{3+}$), and host-intrinsic ultrafast relaxation pathways.

The undoped SrLuF@SrLuF reference sample shows no pronounced emission features at room temperature for wavelength range of 250-750nm, confirming that the observed scintillation originates exclusively from dopant-centered processes. Notably, both $Ce^{3+}$ and $Pr^{3+}$ emissions intensify substantially above ~15% dopant concentration, contrary to bulk scintillators, where concentration quenching typically appears at sub-percent levels. This difference reflects the substitutional nature of $Ce^{3+}/Pr^{3+}$ incorporation on $Lu^{3+}$ sites and the nanocrystalline host environment, which suggests reduced long-range cross-relaxation pathways that normally results in quenching in bulk crystals. The small fluctuations in emission intensity at higher dopant concentrations likely originate from nanoscale thickness inhomogeneities across the drop-casted films, along with slight variations in nanoparticle yield between syntheses.

The mixed spectral character of the $Pr^{3+}$-doped samples, where a broad UV band coexists with sharp 4f-4f lines, offers direct insight into the underlying excited-state pathways. In fluoride hosts, the radiative competition between interconfigurational $4f^1 5d^1 \rightarrow 4f^2$ transitions and intraconfigurational $^1S_0 \rightarrow 4f^2$ / $^3P\_J \rightarrow 4f^2$ channels is dictated by the relative energetic alignment of the crystal-field–split $4f^1 5d^1$ levels and the $^1S_0$ states.[32] Strong crystal fields stabilize the lowest $4f^1 5d^1$ states below $^1S_0$, producing predominantly broad d-f emission, whereas weak-field hosts place $^1S_0$ below the 5d levels, funneling excitation into sharp line emission and enabling photon-cascade processes. In other words, when the lowest 5d state lies above $^1S_0$, relaxation funnels into the $^1S_0$ state, yielding sharp line emission and permitting photon-cascade emission (PCE). In contrast, when the 5d manifold lies below $^1S_0$, broadband 5d→4f emission dominates.[30,33] We hypothesize that the SrLuF nanoscintillators exhibit an intermediate crystal-field regime based on the simultaneous presence of broad and narrow band emissions. This co-existence can indicate that excited electrons initially populate 5d-like states that can either decay radiatively (yielding the broad band) or relax nonradiatively into the $^1S_0$/$^3P\_J$ states, producing the observed sharp f-f transitions consistent with reports in other fluoride materials where small energy separations between the $^1S_0$ level and the lowest 5d states allow dual-channel relaxation. In other words, the $Pr^{3+}$ in SrLuF nanoscintillators experiences a crystal field sufficiently strong to activate 5d-mediated pathways, yet sufficiently weak to leave the $^1S_0$ state energetically competitive.

The time-resolved XEOL measurements (Figure 3c,d) provide insights into the excited-state decay dynamics. A biexponential decay is observed for both $Ce^{3+}$ and $Pr^{3+}$ doped samples, with a sub-1 ns and a sub-15 ns component. The slower decay components in doped samples, while ultrafast, are originating from the f-d transitions in $Ce^{3+}$ and in $Pr^{3+}$. In addition to the sub-1ns decay component, the undoped SrLuF nanoscintillators exhibit a slower decay of 40 ns as well, which is likely associated with defect-mediated radiative recombination.[34,35] Biexponential fits of the decay curves confirm the coexistence of multiple radiative pathways. The full set of fitted lifetimes with weighted decay components plotted in Figures S8 and the extracted decay constants are summarized in S9. The sub-ns decay component is consistent with contributions from cross-luminescence–like processes and has been previously observed in other wide-gap fluorides with possible emission wavelengths in the 190-220 nm range. [15,36] It is worth noting that the decay time of the sub-nanosecond component approaches the temporal resolution limit of the luminescence setup.[36] In materials such as $BaF_2$ and $BaLuF_5$, this emission arises from cross-luminescence, a core-to-valence radiative transition (e.g., Ba 5p → F 2p) that produces sub-nanosecond decay and broadband UV emission. Although SrLuF lacks the heavy alkaline-earth cations that strongly favor classical cross-luminescence, the presence of $Lu^{3+}$, with deep 5p states near the valence band, can support analogous fast core-hole recombination channels. Similar ultrafast components have been reported in Lu-containing fluoride hosts and are often activated or enhanced under high-energy X-ray excitation.[37] We note that direct spectral identification of VUV emission is beyond the scope of the present measurements, and the assignment is based on the observed sub-nanosecond component in conjunction with the wide-bandgap fluoride host. The appearance of this sub-ns decay across all compositions therefore suggests that SrLuF supports a host-intrinsic ultrafast relaxation channel, which persists independently of dopant type or concentration. This host-driven channel operates in parallel with the dopant-centered excitation pathways, giving rise to the multi-timescale scintillation dynamics observed in SrLuF based nanoscintillators.

We evaluate ensemble performance of millimeter-scale self-assembled scintillating superlattices under continuous-wave (CW) and femtosecond high-intensity X-ray excitation to probe scintillation behavior across distinct dose-rate regimes. CW irradiation using a copper-anode laboratory X-ray source establishes scintillation efficiency and imaging response under steady-state photon flux conditions representative of conventional scintillator screens for radiographic imaging, assessed through direct scintillation light imaging. In contrast, femtosecond pulsed excitation deposits energy within ultrashort temporal windows, generating elevated instantaneous dose rates that test scintillation yield, linearity, and radiation stability under extreme irradiation conditions. This dual-regime evaluation constrains scintillator performance across both operational and high-intensity irradiation environments.

Figure 4a-d shows the intensity profile across SrPrF and SrCeF self-assembled crystals upon 30s exposure of CW X-ray of 160kVp. The reference measurements are taken under X-ray off conditions as well confirming the light-out is originating from scintillation. Further experimental considerations were taken into account to minimize not only the ambient light but also possible visible photons that can be generated from the X-ray source itself (the methods section includes further details on the experimental setup). The Pr doped crystals show ~4x intensity boost in scintillated emission for areas "on" and "off" the sample while this number is ~2x for Ce doped sample. Figure 4f shows a cluster of these thin crystals mounted on Kapton tape and similarly the image is recorded for X-ray on and X-ray off post illumination to identify the afterglow. This number is close to 1% for Pr-doped samples arising from possible long-lived f-f transitions while overall, the background decays quickly after exposure. For radiographic imaging, objects such as a metal pin were placed between the X-ray source and the transparent SrPrF scintillator crystals. Upon X-ray irradiation, the transmitted beam created a visible image on the scintillator screen, which was recorded using a digital camera as shown in Figure 4e. Such direct visualization demonstrates the light-output performance of our MLnF scintillators and serves as a proof-of-concept for X-ray imaging.

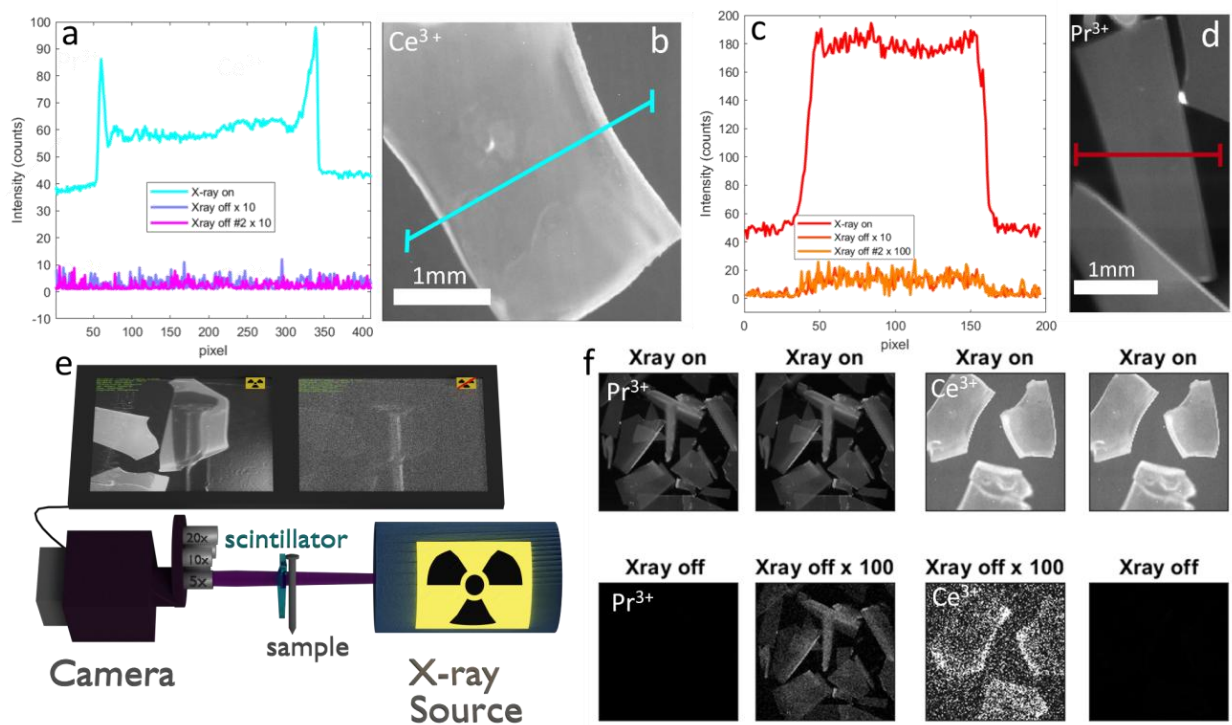

**Figure 4: Scintillation performance and X-ray imaging capabilities of self-assembled SrLuF:RE$^{3+}$ superlattices.**
(a-d) Direct scintillation images of SrPrF and SrCeF mm-scale crystals under continuous-wave 160 kVp X-ray excitation (Cu-anode tube), showing bright, spatially uniform emission. Reference images recorded with the X-ray beam blocked confirm that the signal originates from scintillation.
(e) Radiographic imaging demonstration: a metal pin placed in front of the transparent SrPrF scintillator produces a high-contrast shadow image under X-ray irradiation.
(f) Cluster of thin scintillating crystals mounted on Kapton tape, recorded under X-ray on/off conditions to assess afterglow; Pr-doped samples exhibit <1% residual emission, consistent with weak long-lived 4f–4f contributions.

To probe the scintillation response under extreme irradiation, we subject our mm-scale scintillators to femtosecond, high-intensity X-ray pulses using an X-ray Free Electron Laser (XFEL) at SLAC National Accelerator.[38,39] Figure 5a shows a schematic of the experimental setup. The X-ray pulses are produced by generating short electron bunches accelerating to approximately the speed of light; the self-amplified spontaneous emission causes the electrons to bunch together, generating ultrashort coherent radiation when passed through an undulator. The resulting hard X-ray pulses are among the most brilliant sources of X-rays (high emittance, low divergence, high coherence). In transmission geometry, 9.5 keV X-rays are incident on the scintillator sample at 120Hz, and the scintillation is recorded on a camera on a shot-to-shot basis. Figure 5b shows the average image of the sample under the ambient light, in the dark, and upon X-ray exposure. The radioluminescence (RL) appears where the intense ~50fs, ~20μm focused X-rays illuminate the sample and the images are normalized to the incoming pulse intensity and averaged. When plotting the integrated RL against the incoming X-ray intensity on a shot-to-shot basis the scintillation efficiency corresponds to the slope of the resulting scatter plot. Quantitative analysis of the scintillation efficiency, defined as the ratio of generated photons to the absorbed ionizing energy, reveals that the SrLuF:100%Pr$^{3+}$ scintillator produces approximately 10% of the light output of a commercial YAG:Ce$^{3+}$ scintillator. However the main drawback of YAG:Ce$^{3+}$ scintillators, particularly in ultrafast measurements, is the slower decay lifetime (~150ns)[39] limiting the time-resolution and limiting the applicability at higher repetition rates. While the light output remains below that of state-of-the-art bulk scintillators, the combination of ultrafast decay, radiation hardness, and scalable superlattice assembly establishes a distinct performance regime, with potential pathways for improving light yield through further compositional and photonic optimization (nanophotonic structuring). We note that absolute photon yield (photons/MeV) was not calibrated in this study; instead, relative light output is reported against YAG:Ce$^{3+}$ under identical excitation conditions to enable direct comparison of emission dynamics and intensity trends.

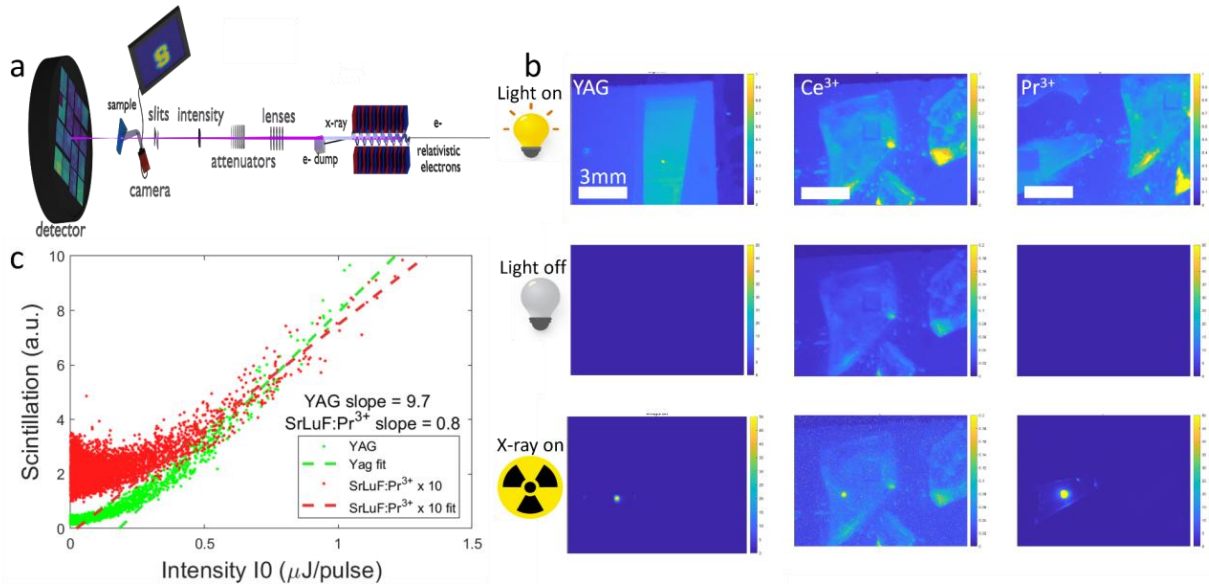

**Figure 5: Ultrafast scintillation of SrLuF:RE$^{3+}$ scintillator assemblies under femtosecond XFEL excitation.**
(a) Schematic of the XFEL measurement, where 9.5 keV, 120 Hz femtosecond X-ray pulses are focused to ~50 μm and transmitted through the scintillator, with radioluminescence recorded on a shot-to-shot basis.
(b) Average images of the SrLuF:Pr scintillator under ambient illumination, dark conditions, and XFEL excitation, normalized to the incident pulse intensity.
(c) Integrated radioluminescence plotted against incoming pulse energy, showing linear response and enabling extraction of scintillation efficiency.
(d) Relative scintillation efficiency of SrLuF:Pr$^{3+}$ compared to commercial YAG:Ce$^{3+}$, demonstrating that the MLnF platform produces ~10% of the YAG:Ce$^{3+}$ light output under identical XFEL conditions.

To evaluate the radiation hardness and damage threshold, the most promising SrLuF:100%Pr$^{3+}$ sample was evaluated under 5 different beam conditions (incoming pulse energy) for an extended time of ~16000 shots or approximately 2 minutes at 120 Hz. Radiation hardness is a critical yet often underexplored property in the context of nanoscintillators. Figure 6a shows the RL intensity vs. exposure time in seconds (where each dot is an individual 50fs X-ray pulse). From top to bottom the incident pulse energy varies between 0.1µJ/pulse and 40µJ/pulse. The black vertical lines indicate moving the sample to a fresh location on the sample to further evaluate sample homogeneity and reset potential damage. Figure 6b shows the RL plotted against the incoming X-ray intensity with linear correlation describing an ideal scintillator. For 0.1-1 µJ/pulse remains linear and shows a proportional scintillation response throughout but for 4-40 µJ/pulse the RL exhibits decay during the exposure time indicating damage. Thus, the SrLuF:Pr sample is well suited for intense X-ray irradiation as long as the pulse energy remains below ~5mJ/mm$^2$ per shot (9.5keV/1µJ/50fs/20µm/1eV BW) corresponding to a peak intensity of $10^{13}$ W/cm$^2$. Similar single-shot thresholds have been reported for commercial YAG:Ce$^{3+}$ at 480 mJ/mm$^2$ and for a Gd$_2$O$_2$S:Pr ceramic at 13.5mJ/mm$^2$.[40]

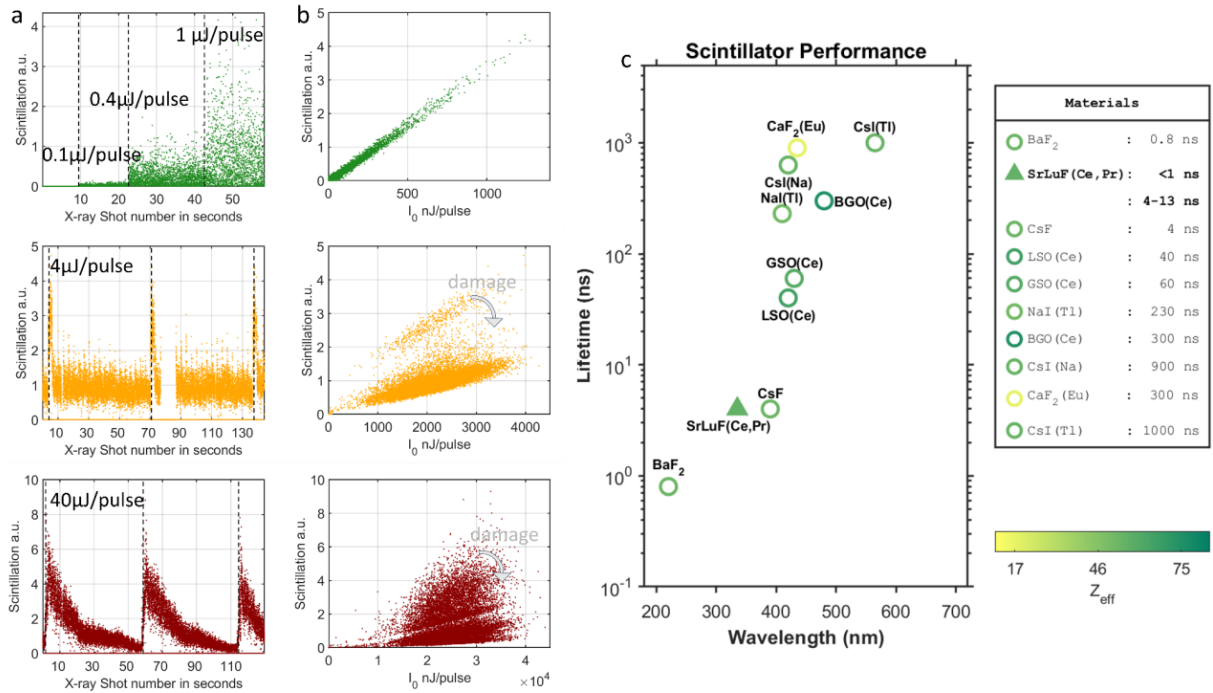

**Figure 6: Radiation hardness, proportional response, and comparative scintillator performance of MLnF superlattice scintillators under extreme X-ray irradiation.**
(a) Radioluminescence (RL) intensity of the SrLuF:100%Pr superlattice scintillator as a function of exposure time under femtosecond XFEL irradiation for incident pulse energies from 0.1 to 40 µJ per pulse (50 fs, 120 Hz). Vertical lines indicate translation to a fresh sample region.
(b) Shot-to-shot RL intensity as a function of incoming X-ray pulse energy. Linear scaling indicates proportional scintillation response for pulse energies below 1µJ; deviations at high pulse energies (4-40 µJ per pulse) arise from radiation-induced damage.
(c) Comparative scintillator performance map showing decay lifetime versus emission wavelength for representative inorganic scintillators (BaF$_2$, NaI:Tl, CsI:Tl) and the MLnF scintillators developed in this work.

Finally, we map the MLnF-based nanoscintillator platform onto the broader commercial inorganic scintillator landscape by comparing effective atomic number, decay lifetime, and emission wavelength (Figure 6c). Figure 6c presents a comparative performance map highlighting the relationship between scintillation decay constant and emission wavelength across widely used scintillators. Reference scintillators include $BaF_2$, which exhibits sub-nanosecond decay with deep-UV emission (~220 nm), and thallium-doped alkali halides NaI:Tl and CsI:Tl, two of the most widely deployed scintillators, with characteristic decay lifetimes of approximately 200 ns and 1000 ns, respectively. In this landscape, the MLnF superlattice scintillators occupy a distinct regime that combines high effective atomic number ($Z_{eff} \approx 54.5$), near-visible emission (310-335 nm), and ultrafast timing (sub-nanosecond to sub-15 ns), positioning them between ultrafast fluoride scintillators and high-light-yield alkali halides. This regime is particularly attractive for applications requiring both fast timing and high stopping power, such as high-rate X-ray and γ-ray detection, time-of-flight imaging, and ultrafast photon science, where conventional scintillators typically trade light yield against timing performance.

**Conclusion and outlook**

We have demonstrated a bulk scintillator architecture constructed from nanoscale radioluminescent building blocks, showing that near-atomic control over composition, dopant distribution, and interfacial structure can be translated into macroscopic materials with high stopping power, ultrafast timing, and radiation robustness. SrLuF:$Ce^{3+}$/$Pr^{3+}$@SrLuF core-shell nanoscintillators assembled into millimeter-scale superlattices combine a high effective atomic number, environmental stability, sub-nanosecond and single-digit nanosecond decay dynamics, proportional response, and optical yields within an order of magnitude of commercial YAG:$Ce^{3+}$. The core-shell structure suppresses defect-mediated quenching and Frenkel disorder while enabling efficient dipole- and spin-allowed 4f-5d emission from $Ce^{3+}$ and $Pr^{3+}$ activators. The nanocrystalline host environment further permits unusually high dopant concentrations without the concentration quenching that limits bulk scintillators, providing an independent tuning knob over brightness and temporal response. Looking forward, this platform opens a new materials space for radiation detection and ultrafast photon science. On the materials side, further optimization of nanocube size, shell thickness, and dopant distributions will enable deterministic control over stopping power, radiative efficiency, and decay kinetics, while extension to heavier alkaline-earth and lanthanide fluoride hosts provides a direct pathway toward efficient γ-ray detection for nuclear security, astrophysics, and high-energy photon science, where fast timing, radiation hardness, and proportionality are essential. In parallel, integration of these nanoscintillators with emerging nanophotonic architectures offers a route toward next-generation "super-scintillators" with engineered emission extraction, spatiotemporal confinement, and GHz-rate response.

Beyond radiation detection, the modularity, stability, and nanoscale processability of MLnF nanoscintillators open new opportunities in radiation-enabled nanomedicine. In particular, their efficient ionizing-to-optical energy conversion and tunable emission spectra provide a foundation for X-ray activated photodynamic therapy and image-guided theranostics, where deeply penetrating radiation can be converted locally into visible light to activate photosensitizers and generate cytotoxic reactive oxygen species. More broadly, this work illustrates how bottom-up materials design, when coupled with scalable colloidal synthesis and self-assembly, can transform nanoscale photophysics into macroscopic functionality, enabling a new generation of radiation-responsive materials whose performance is encoded at the level of individual nanocrystals and expressed at device scale.

**Methods**

**Synthesis of MLnF Core-Shell Nanoscintillators**

The alkaline-earth rare-earth fluoride $M_{1-x}Ln_xF_{2+x}$ (MLnF) core nanoparticles are synthesized via thermal decomposition of the corresponding trifluoroacetic acid (TFA) salts in the presence of capping ligands in which the $Ln^{3+}$ salts are decomposed at temperatures around 300 °C with minor modification to the protocol reported by Fischer et al.[28] and Siefe et al.[29] To increase the luminescence efficiency the core nanoparticles are shelled in a separate step. The shell layer suppresses the potential detrimental impact of surface defects on active core particles as major energy loss contributors and other potential factors that may result in quenching the luminescence emissions. During a post processing procedure, the pre-prepared trifluoroacetate shell precursor is hot injected (~270 °C) into the core nanoparticle solution via a syringe pump with slow injection rate (2 mL/hr) and incubated to assure efficient nucleation and growth of the shell. The final core-shell products are capped with oleate ligands.

**Chemicals and Materials**

Strontium carbonate (99.9%, Sigma-Aldrich) and rare-earth oxides (99.9%, Alfa Aesar), Trifluoroacetic acid (99%, Alfa Aesar), cerium acetate (99.99%) and oleic acid (90%, Sigma-Aldrich) are used for this synthesis. Oleylamine (70% technical grade) and 1-octadecene (90% technical grade) are obtained from Acros, and additional organic solvents are purchased from Fisher Scientific. All reagents are used without further purification.

**Preparation of Metal Trifluoroacetate Precursors**

Precursor synthesis is based on established protocols with minor modifications to accommodate variations in the solubility of lanthanide oxides and alkaline-earth carbonates. A typical procedure for praseodymium involved dissolving 1 mmol $Pr_2O_3$ in a mixture of 1 mL of 99% trifluoroacetic acid and 5 mL of deionized water in a three-neck flask. The solution is maintained at 90°C in an oil bath using a condenser to prevent evaporation until complete dissolution is achieved, yielding a clear solution. Subsequent evaporation at 65°C is removed residual water and acid, affording a white powder corresponding to 2 mmol of the praseodymium trifluoroacetate complex. For other metal precursors, the relative amounts of water and acid are adjusted, typically requiring less reagent and initiating dissolution at ambient conditions; however, the solution is still heated to 90°C to ensure complete conversion.

**Synthesis of $M_{1-x}Ln_xF_{2+x}$ Core Nanoscintillators**

Core nanoparticles are synthesized via thermal decomposition of the corresponding trifluoroacetate salts. Equimolar solutions (1 mmol each) of the lanthanide and strontium trifluoroacetates were combined in a 50 mL three-neck flask under magnetic stirring (approximately 300 rpm). To this mixture, oleic acid, 1-octadecene, and oleylamine are added to reach final volumes of 6 mL, 13 mL, and 2 mL, respectively. The addition of oleylamine are induced an approximate 6°C increase in temperature, and the stirring rate is subsequently increased to 1000 rpm. The reaction mixture is degassed under vacuum (cycling between vacuum and argon 3 times) at 120°C until residual water and oxygen are removed. Upon achieving a pressure below 40 mTorr, the temperature is ramped to 300°C over 10-15 minutes under an argon atmosphere. Rapid decomposition of the trifluoroacetate salts between 270°C and 300°C results in an exothermic eruption and the evolution of water vapor, which is mitigated via continuous argon flow through a bubbler or by puncturing the septum. After maintaining the reaction at 300°C for 60 minutes, the system is cooled to room temperature, and the particles are isolated by ethanol precipitation and centrifugation (approximately 5000 g). A washing cycle involving hexane dispersion and ethanol precipitation (centrifugation at roughly 4000 g) yield core nanoscintillators, which are stored in hexane.

**Shell Growth Procedure**

For shelling, 0.75 mmol of core nanoparticles are combined in a 100 mL three-neck flask with 7.5 mL oleic acid and 15 mL 1-octadecene under moderate stirring. Following hexane removal at 65-75 °C (with the flask necks open), the system is degassed under vacuum at 120 °C under magnetic stirring and subsequently maintained at this temperature. After three vacuum/argon cycles to eliminate residual gases, the temperature is raised to 310 °C under an inert argon atmosphere. At 270 °C, a shell precursor solution, is prepared by mixing Lu-oleate and Sr trifluoroacetate in a 1:1 molar ratio followed by degassing under vacuum, is injected dropwise via a syringe pump at a rate calibrated to approximately 6 mmol precursor per mmol of core material, corresponding to an injection rate of 2 mL hr$^{-1}$ based on the core concentration. The final product is washed by ethanol-induced precipitation, centrifugation at ~4000 g, and three washing cycles with hexane and ethanol before final dispersion in cyclohexane.

**X-ray Scintillation Measurements**

**Steady-State XEOL Measurements**

Steady-state X-ray excited optical luminescence of MLnF nanoparticles is measured at room temperature using a Bruker FR591 rotating anode X-ray tube operated at 50 kVp and 50 mA with a copper anode. The X-ray beam is directed perpendicularly onto the sample, and the integrated luminescence spectrum is collected using a SpectraPro-2150i spectrometer coupled to a CCD detector (PIXIS:100B) with a 500 μm mono-slit. Emission spectra are corrected for spectrometer transmission, detector sensitivity, and dark current to ensure accurate room-temperature radioluminescence data.[34,41]

**Time-Resolved XEOL Measurements**
Time-resolved XEOL is characterized by using a custom-built pulsed X-ray system employing time-correlated single photon counting (TCSPC). The system consists of a light-excited X-ray tube (Hamamatsu N5084) operated at 40 kVp and triggered by an ultrafast Ti:sapphire laser (200 fs pulses, Coherent Mira, operating at 165 kHz). Emission from the samples is detected without wavelength discrimination by a Hamamatsu multichannel plate photomultiplier (R3809U-50) biased at 3000 V, and the signal is processed through an Ortec 9308 ps analyzer. The overall instrument response function (IRF) is determined to have a full width at half maximum (FWHM) of ~100 ps.[34,41]

Decay profiles are analyzed by fitting a bi-exponential decay model to the data. The fitting function comprises two exponential terms, with amplitudes "$a_1$" and "$a_2$" and corresponding lifetimes "$\tau_1$" and "$\tau_2$", along with a constant background "C". This sum of exponentials is convolved with an error function broadened by a Gaussian "irf" (in FWHM) to account for instrumental broadening.

$$C + 0.5 \cdot a_1 \cdot \exp\left(\frac{\mathrm{irf}^2 - 2(x - tzero)\cdot\tau_1}{2\cdot\tau_1^2}\right) \cdot \left(1 - \mathrm{erf}\left(\frac{\mathrm{irf}^2 - (x - mp)\cdot\tau_1}{\sqrt{2}\cdot\mathrm{irf}\cdot\tau_1}\right)\right)$$
$$+ 0.5 \cdot a_2 \cdot \exp\left(\frac{\mathrm{irf}^2 - 2(x - tzero)\cdot\tau_2}{2\cdot\tau_2^2}\right) \cdot \left(1 - \mathrm{erf}\left(\frac{\mathrm{irf}^2 - (x - tzero)\cdot\tau_2}{\sqrt{2}\cdot\mathrm{irf}\cdot\tau_2}\right)\right)$$

*Erf broadened by gaussian followed by biexponential decay

**Scintillation Light-imaging: Scintillation Light Imaging (XCT):** Scintillation imaging is performed using a Zeiss Xradia 520 Versa X-ray CT system equipped with a polychromatic X-ray source operated at 160 kV, 10 W, and 65 μA with a 30 s exposure time. A 5X objective lens is used in combination with a superlattice scintillating crystal to visualize scintillation light. The system employs X-ray tubes with copper targets and offers a tube voltage range of 30–160 kV.

**Radiation Hardness and Linear Response Measurements:** Radiation hardness measurements are conducted at the XCS beamline at LCLS. ~50fs, 9.5keV X-ray pulses are focussed down to 20 μm spot size using compound refractive beryllium lenses. The radioluminescence resulting from each X-ray pulse is captured on an Alvium 1800u-240 camera at 120Hz. The single shot diffraction patterns are recorded on an ePix10k2M single shot detector placed downstream of the sample to evaluate the sample crystallinity and monitor for shot-to-shot structural degradation. The sample is mounted in air, on an XYZ stack of XA05A-L202-R Kohzu linear stages, allowing for raster scanning of the crystal samples at the interaction point. The incident pulse energy is attenuated using a series of solid attenuators and the pulse energy of each shot determined by the intensity-position monitors (IPMs) installed at the beamline[42].

**Transmission Electron Microscopy (TEM):** TEM images are acquired using a FEI Tecnai G2 F20 X-TWIN microscope operated at an accelerating voltage of 200 kV. Samples are prepared by drop-casting approximately 5 μL of a dilute dispersion of the nanoparticles in hexanes onto ultrathin carbon type-A, 400 mesh copper grids (Ted Pella, Inc.), followed by solvent evaporation at room temperature.

**Inductively Coupled Plasma Optical Emission Spectroscopy (ICP-OES):** The lanthanide composition of the core UCNPs is quantified using inductively coupled plasma optical emission spectroscopy (ICP-OES). Approximately 5 mg of each sample, assuming 50% organic content, is dissolved in 3% aqueous nitric acid (prepared with ICP-grade reagents) at a concentration of 10 mg/mL. Calibration standards for each element are prepared from Ultra Scientific ICP standards. Elemental concentrations are measured in ppm and converted to relative molar percentages. $Pr^{3+}$ and $Ce^{3+}$ levels in doped- SrLuF core particles are found to be near the targeted ratios, while $Ce^{3+}$-based samples exhibited lower-than-expected doping levels for intermediate doping levels.

**X-ray Photoelectron Spectroscopy (XPS):** XPS measurements are performed using a PHI VersaProbe-III spectrometer equipped with a monochromatic Al-Kα X-ray source (1486.6 eV). The system allows for adjustable X-ray spot sizes ranging from 10 μm to 200 μm, enabling high-resolution surface characterization. Survey scans are initially acquired to identify elemental composition, followed by high-resolution scans of the Ce 3d and Pr 3d regions to analyze the oxidation states of cerium and praseodymium on SrPrF and SrCeF core nanoparticles. The nanoparticles are dropcasted on silicon substrate. Charge neutralization is applied during measurements to mitigate surface charging effects. Spectral analysis is done using MultiPak software.

**X-ray Diffraction (XRD):** Powder X-ray diffraction (XRD) is used to determine the crystal structure of the MLnF nanoscintillators. XRD patterns are collected over a 2θ range of 10-90° using Cu Kα$_1$ radiation (λ = 0.15406 nm) at 40 kV and 45 mA, with a step size of 0.05° and a scan rate of 4°/min. The reference pattern for SrLuF is obtained from the International Center for Diffraction Data (ICDD). All compositions are confirmed to maintain an FCC close-packed structure (Fm-3m space group), with a lattice parameter of a = 5.7 Å.

**Acknowledgments**
P.M. acknowledge the support from the Stanford Bio-X seed grant under award no. 1248231-100-WXDCW and Stanford Molecular Imaging Scholars Program (SMIS) under award no. NIH T32 CA118681 as well as NIH R21 EB033551. Additionally, P.M. and J.A.D. acknowledge support from the Q-next grant under award no. DE-AC02-76SF00515. Q.F. and A.M.L. acknowledge support from the Department of Energy, Office of Basic Energy Sciences, Division of Materials

Science and Engineering, under contract DE-AC02-76SF00515. A.S. acknowledges support from the National Science Foundation (NSF) through the Graduate Research Fellowships Program. Use of the Linac Coherent Light Source (LCLS), SLAC National Accelerator Laboratory, is supported by the U.S. Department of Energy, Office of Science, Office of Basic Energy Sciences under Contract No. DE-AC02-76SF00515. The authors also acknowledge insightful and fruitful discussions with Dr. Charles Roques-Carmes.

AUTHOR INFORMATION
**Corresponding Authors**
Parivash Moradifar (pmoradi@stanford.edu), Craig S. Levin (cslevin@stanford.edu), Jennifer Dionne (jdionne@stanford.edu).

**Author Contributions**
P.M. designed the study along with J.D., C.L. and G.C. P.M. carried out the synthesis, sample preparation, structural analysis, XEOL, X-ray imaging as well as data analysis of all presented data. T.B.v.D. designed the XEOL, X-ray imaging and XFEL experiments with P.M. and supervised the data analysis. T.B.v.D. performed the XFEL experiments. P.M. and T.B.v.D. contributed equally to this work. Q.F., A.L. carried out time-resolved experiments. J.D. and C.L. supervised the work and obtained research funding. The manuscript was drafted by P.M. and edited by all co-authors. M.F., C.S., A.S., contributed to the synthesis. F.M., D. J. contributed to XEOL measurements. A.L. and Q.F. helped design and carry out the ultrafast scintillation lifetime measurements.